\documentclass[twocolumn,showpacs,superscriptaddress,aps,prb,floatfix,10pt]{revtex4-2}
\usepackage{graphicx}
\usepackage{amsmath,amssymb}
\usepackage[utf8]{inputenc}
\usepackage{natbib}
\usepackage[bookmarks=false]{hyperref}
\usepackage{xcolor}
\hypersetup{colorlinks=true,citecolor={blue},linkcolor={blue},urlcolor={blue}}
\usepackage{units}
\usepackage[normalem]{ulem}
\usepackage{wasysym}
\usepackage{soul}
\usepackage{adjustbox}
\newcommand{\rom}[1]{\uppercase\expandafter{\romannumeral #1\relax}}

\makeindex

\begin{document}

	\definecolor{lime}{HTML}{A6CE39}
	
	\definecolor{orcidlogocol}{HTML}{A6CE39}
	
	\definecolor{lilac}{RGB}{150,100,180}
	\setstcolor{red}
	\newcommand{\Stella}[1]{{\color{lilac}{#1}}}
	\newcommand{\hs}[1]{\textcolor{purple}{#1}}
	
	\newcommand\hcancel[2][red]{\setbox0=\hbox{$#2$}%
		\rlap{\raisebox{.45\ht0}{\textcolor{#1}{\rule{\wd0}{1pt}}}}#2}

	\title{Minor embedding with Stuart-Landau oscillator networks}

	\date{\today}
	
	\author{S. L. Harrison}
	\email{S.L.Harrison@soton.ac.uk}
	\affiliation{School of Physics and  Astronomy, University of Southampton, Southampton, SO17 1BJ, United Kingdom}
	
	\author{H. Sigurdsson}
	\affiliation{Science Institute, University of Iceland, Dunhagi 3, IS-107, Reykjavik, Iceland}
	\affiliation{School of Physics and  Astronomy, University of Southampton, Southampton, SO17 1BJ, United Kingdom}
	
	\author{P. G. Lagoudakis}
	\affiliation{Hybrid Photonics Laboratory, Skolkovo Institute of Science and Technology, Territory of Innovation Center Skolkovo, 6 Bolshoy Boulevard 30, Bld. 1, 121205 Moscow, Russia}
	\affiliation{School of Physics and  Astronomy, University of Southampton, Southampton, SO17 1BJ, United Kingdom}

\begin{abstract}
We theoretically implement a strategy from quantum computation architectures to simulate Stuart-Landau oscillator dynamics in all-to-all connected networks, also referred to as complete graphs. The technique builds upon the triad structure minor embedding which expands dense graphs of interconnected elements into sparse ones which can potentially be realized in future on-chip solid state technologies with tunable edge weights. As a case study, we reveal that the minor embedding procedure allows simulating the XY model on complete graphs, thus bypassing a severe geometric constraint.
\end{abstract}
	
\pacs{}
\maketitle

\section{Introduction}
Solving dense graph combinatorial optimization problems has been studied thoroughly in the field of quantum annealing, where qubits and their relative coupling strengths represent graph vertices and edge weights respectively. To solve arbitrary graph problems using pre-existing qubit architectures, dense graphs can be embedded into new representative graphs using gauge field constraints~\cite{lechner_quantum_nodate, puri_quantum_2017}, or connected ferromagnetic subgraphs using \textit{minor embedding} techniques~\cite{choi_minor-embedding_2011,cai_practical_2014,king_algorithm_2014,venturelli_quantum_2015,perdomo-ortiz_readiness_2019,hamerly_experimental_2019}. In the latter, a standardized minor embedding to the \textit{triad structure} is commonly used, allowing any all-to-all connected (dense) graph to be mapped to a planar qubit architecture \cite{choi_minor-embedding_2008,choi_minor-embedding_2011,boothby_fast_2016}. 
	
In this study, we investigate the feasibility of using minor embedding to a triad structure~\cite{choi_minor-embedding_2011,boothby_fast_2016} to simulate the dynamics in dense networks of interacting classical oscillators instead of qubits. Our study is motivated by the recent developments in designing analogue computing strategies based on optical oscillatory networks to heuristically solve complex graph problems across various platforms such as: Exciton-polariton condensates~\cite{berloff_realizing_2017,Kyriienko_PRB2019, harrison_solving_2020,luo_classical_2020, Kavokin_NatRevPhys2022, Tao_NatMat2022}, photon condensates \cite{vretenar_controllable_2021}, non-degenerate optical parametric oscillators~\cite{tamate2016simulating, Takeda_QST2017,reifenstein_coherent_2021}, photon down-conversion oscillators~\cite{Latifpour_CommPhys2022}, and coupled microlaser arrays \cite{Gershenzon_NanoPho2020, reddy_phase-locking_2021}. Even digital computer algorithms designed to efficiently solve the dynamics of oscillatory networks have displayed impressive prowess in combinatorial optimization~\cite{Goto_SciAdv2019, Bohm_CommPhys2021}.

Optical platforms for unconventional analogue computation~\cite{Latifpour_CommPhys2022} have many desirable properties such as photonic parallelism, low cross-talk, ultrafast timescales (i.e., high sampling rate), and low power consumption using passive elements. However, engineering high degrees of graph connectivity---which has been achieved with good control in cold atom ensembles in optical cavities~\cite{Periwal_Nature2021}---presents still a major hurdle in many of the abovementioned photonic platforms. As an example, in planar microcavity exciton-polariton condensates the coupling strength between neighboring condensates decays exponentially with their spatial separation distance~\cite{Topfer_CommPhys2020, harrison_synchronization_2020}, making the coupling beyond nearest neighbors usually negligible. That, plus the need to control each inter-condensate coupling strength makes it nigh on impossible to solve an all-to-all connected graph beyond a handful of condensate vertices. A scheme to achieve all-to-all coupling between polariton condensate modes was theoretically proposed using a fast time-modulated nonresonant laser~\cite{Sigurdsson_ACSPho2019}. We show that these connectivity problems can, in principle, be overcome using the minor embedding technique to the triad structure, which realizes a more experimentally friendly, on chip, strategy to simulate dynamics and emergent behaviours in all-to-all connected networks of nonlinear optical oscillators.

We focus on two measures, and compare them between non-embedded and corresponding embedded networks, to quantify the applicability of our technique. Firstly, calculate and characterize the emergence of coherence in disordered ferromagnetically connected Stuart-Landau oscillatory networks~\cite{acebron_kuramoto_2005} with increasing coupling strength. Second, we investigate the efficiency of embedded oscillatory networks to approximate low energy solutions of the classical XY Hamiltonian through dynamical annealing~\cite{Leleu_PRE2017, Kalinin_NJP2018}.

	
	\section{The Stuart-Landau Model}
The Stuart-Landau model describes a plethora of oscillatory systems and is formally derived from the normal form of an Andronov-Hopf bifurcation. We apply this universal model to describe the dynamics of our dissipatively coupled oscillators at the graph vertices written:
	\begin{equation}
	\frac{d\psi_n}{dt} = -\left[ i\omega_n +|\psi_n|^2\right]\psi_n + \sum_{m=1}^N J_{n,m} \psi_m
	\label{eq:SL}
	\end{equation}
	Here, $\psi_n \in \mathbb{C}$ denotes the $n$th oscillator, $\omega_n$ its intrinsic frequency, and $J_{n,m}$ is the coupling strength between oscillator $n$ and $m$ corresponding to the weighted edge connecting graph vertices $n$ and $m$. We can re-write the Stuart-Landau model in polar coordinates with $\psi_n = \rho_n e^{i\theta_n}$ arriving at:
	\begin{align}
	\frac{d\rho_n}{dt} & = -\rho_n^3 + \sum_m J_{n,m}\rho_m \cos(\theta_m - \theta_n); \label{eq:RE} \\
	\frac{d \theta_n}{dt} & = \omega_n + \sum_m J_{n,m} \frac{\rho_m}{\rho_n} \sin(\theta_m - \theta_n). \label{eq:IM} 
	\end{align}
	If the amplitudes of the oscillators are set as equal $\rho_n = \rho_0$, and with zero detuning $\omega_n=0$, then \autoref{eq:IM} describes the generalized \textit{Kuramoto model} \cite{kuramoto_self-entrainment_1975,strogatz_kuramoto_2000}. The fixed points of a Kuramoto network are described by the minima of the following Lyapunov potential~\cite{acebron_kuramoto_2005},
	\begin{equation}
\mathcal{L} = -\sum_{n,m} J_{n,m} \cos{(\theta_n - \theta_m)},
\end{equation}
which is the same as the well known XY Hamiltonian. Therefore, the phases $\theta_n$ in a Kuramoto network experience a gradient descent towards local minima of the XY Hamiltonian $\dot{\theta}_n = - \partial \mathcal{L} / \partial \theta_n$. In this respect, the phasors $\mathbf{s}_n = [\cos{(\theta_n)}, \sin{(\theta_n)}]^\text{T}$ play the role of two-dimensional classical spins which experience phase space flow towards these minima. This is in a similar spirit to Ising machines which undergo quantum or classical annealing into the ground state of some target Ising Hamiltonian~\cite{Mohseni_NatRevPhys2022}. In this sense, $J_{n,m}>0$ are said to be ferromagnetic links (in-phase coupling) and $J_{n,m}<0$ are antiferromagnetic links (anti-phase coupling). 

Surprisingly, even with coupling induced amplitude inhomogeneity $\rho_n \neq \rho_m$ Stuart-Landau networks have displayed impressive results in approximating the global minima of the XY Hamiltonian in dense networks~\cite{Kalinin_NJP2018}. In Sec.~\ref{sec.XY} we will characterize the quality of using a minor embedding technique to simulate densely connected Stuart-Landau networks by comparing their performance in approximating low energy solutions of the XY Hamiltonian against corresponding non-embedded networks. 

	\section{Minor Embedding}
	\subsection{Triad Graph}
\begin{figure}[b]
		\centering
		\includegraphics[width=0.8\linewidth]{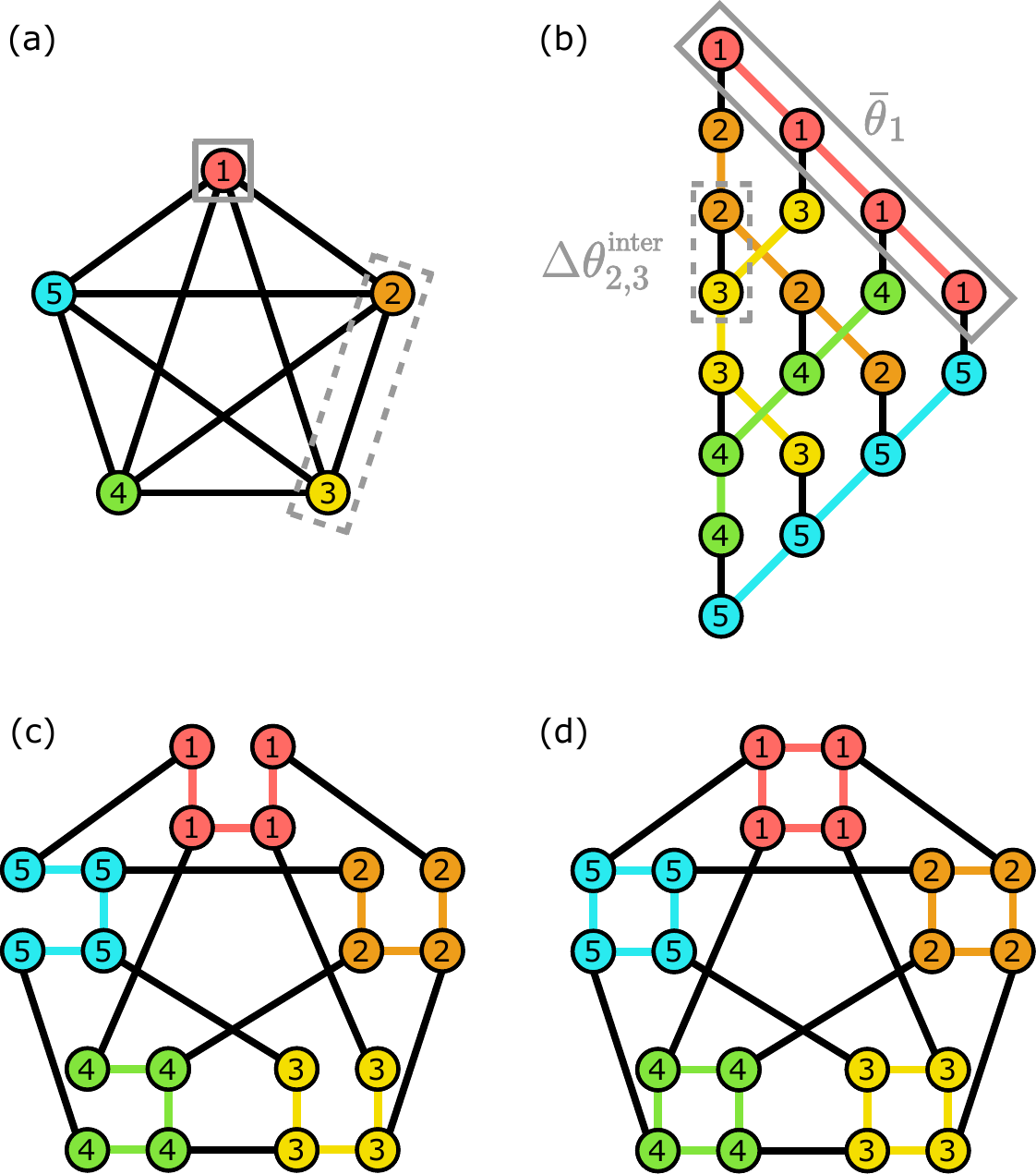}
		\caption{(a) $K_5$ graph mapped through minor embedding to (b) the $K^\text{emb}_5$ triad graph, (c) the mathematical layout of $K^\text{emb}_5$ and (d) when including $N$ additional edges to loop the chains. In (b-d), the FM chain couplings are shown by colored edges $J^\text{intra}=J_c$ and the encoded coupling strength of $K_5$ are shown by black edges, such that $J^\text{inter}_{n,m} = J_{n,m}$. The grey solid and dashed boxes overlayed on (a,b) demonstrate two possible phase extraction methods for the unembedded and embedded XY energies from the triad graph respectively.}
		\label{fig:triad_schem}
	\end{figure}

	In the process of creating a triad graph, we must first consider the undirected complete graph (all-to-all connected graph) of $N$ vertices $K_N$, with vertex set: $V(K_N)$ and edge set: $J(K_N)$. Each vertex $V_n$ is assigned an index $n$ and each edge symmetrically connecting two distinct vertices $V_n$ and $V_m$ is denoted $J_{n,m} = J_{m,n}$. Through the process of minor embedding, $K_N$ is mapped to the triad graph $K^\text{emb}_N$ [shown for $K_5$ in \autoref{fig:triad_schem}(a,b)] by expanding each vertex $V_n\in V(K_N)$ to a chain of uniform coupled vertices of length $N-1$, with intra-chain edge weights set to,
	\begin{equation}
	J^\text{intra} = J_c>0, \qquad \text{(colored edges)}. 
	\label{eq.colored_edges}
	\end{equation}
	Each vertex of the chain is adjacent to a single vertex of a another chain with inter-chain edge weights,
	\begin{equation}
	J^\text{inter}_{n,m} = J_{n,m}, \quad \forall \ J_{n,m}\in J(K_N), \ \  \text{(black edges)}. 
	\label{eq.black_edges}
	\end{equation}
	Additional description can be found in~\cite{footnote1}.

Physically, $J_c>0$ gives precedence to in-phase locking between oscillators within each chain, which---in the context of the XY Hamiltonians---can be regarded as a ferromagnetic (FM) type coupling. The aim of FM coupling is to minimize the deviation between oscillators across each chain, in order to better simulate the dynamics of the complete graph $K_N$~\cite{hamerly_experimental_2019}. In other words, the red vertex in the complete graph [red oscillator $\psi_1$ in \autoref{fig:triad_schem}(a)] is represented by the average amplitude and phase of the oscillators in the red chain of the triad [see grey solid box in \autoref{fig:triad_schem}(b)]. Notice that the number of black edges in the complete graph is the same as in the triad graph. 

In this study, we will work close to the bifurcation threshold of the system separating the attenuated state $\rho_n = 0$ from the oscillatory state $\rho_n \neq 0$, and are thus interested only in the phases of the oscillators whose average in each colored chain is written,
	\begin{equation}
	\bar{\theta}_n = \frac{1}{N-1}\sum_{n' \in \text{chain}}^{N-1} \theta_{n,n'}.
	\end{equation}
	We will refer to this phase averaging as {\it unembedding} the triad graph~\cite{pudenz_parameter_2016,pelofske_advanced_2020}. That is, the averaged phases $\bar{\theta}_n$ of the triad graph have been unembedded back to the vertices of the original complete graph. 
	
	Where the triad structure in \autoref{fig:triad_schem}(b) is the hardware layout for investigating the minor embedding of complete graphs, it can be easier to picture the mapping process as the mathematical layout shown in \autoref{fig:triad_schem}(c), where each vertex of $K_N$ is expanded to a chain of length $N-1$ about its original vertex site, such that the adjacency of the hardware layout [\autoref{fig:triad_schem}(b)] and mathematical layout [\autoref{fig:triad_schem}(c)] are equivalent.
	
We also consider the addition of an intra-chain edge between the first and last vertex of each chain [\autoref{fig:triad_schem}(d)], thus ``looping'' them, such that the looped graph has a uniform degree of connectivity. Although this is more difficult to realize experimentally in most platforms, we will compare results between unlooped and looped chains at minimising the XY Hamiltonian, and elucidate on how this more ``symmetric'' connectivity affects the performance of the triad graph.
	
	\subsection{XY Energy Extraction}
	As mentioned above, each vertex of the complete graph $K_N$ is associated with a complex valued number $\psi_n$ containing information on the state the $n$th oscillator. We define the state vector of the graph as $\boldsymbol{\psi} = [\psi_1,\, \psi_2,\, \dots \psi_N]^\text{T}$ and its corresponding phase vector $\boldsymbol{\theta} = [\theta_1,\, \theta_2,\, \dots \theta_N]^\text{T}$. The energy of the graph $K_N$ is calculated from the XY Hamiltonian written,
	\begin{equation}
	H_{XY} = -\sum_{n,m}^N J_{n,m}\cos{(\theta_n - \theta_m)}.
	\label{eq:orig_XY}
	\end{equation}
	On the other hand, the energy of the triad graph can follow two methods. The first one defines the energy of the unembedded triad in the same way as \autoref{eq:orig_XY} but with the average phases across the chains $\bar{\theta}_n$,
	\begin{equation}
	H^\text{unemb}_{XY} = -\sum_{n,m}^N J_{n,m}\cos (\bar{\theta}_{n} - \bar{\theta}_{m}).
	\label{eq:unembedded}
	\end{equation}
We refer to this as the {\it unembedded energy} of the triad. 

The other method directly uses the relative phase $\Delta \theta^\text{inter}_{n,m}$ between all pairs of oscillators still ``embedded'' in the triad graph connected by a black inter-chain edge as depicted by the grey dashed boxes in \autoref{fig:triad_schem}(a,b). We will refer to this as the {\it embedded energy} of the triad graph written,
	\begin{equation}
	H^\text{emb}_{XY} = -\sum_{n,m}^N J_{n,m}\cos{(\Delta \theta^\text{inter}_{n,m})}.
	\label{eq:embedded}
	\end{equation}
	Note that when $J_c =0$ then minimising $H^\text{emb}_{XY}$ is trivial since the graph forms just a set of $N$ individual oscillator pairs connected by $J_{n,m}$.
	

	\section{Results}
\subsection{Coherence Properties}
Here, we numerically investigate and compare the phase coherence properties (i.e., the ability to synchronize in-phase) in the oscillator network dynamics between the complete graph and the triad graph. We consider a complete FM graph where each oscillator is coupled equally to all the others with coupling strength $J_{n,m} = J > 0$. The frequencies $\omega_n$ are randomly chosen from a normal distribution with density $g(\omega)$ of mean $\bar{\omega}$ and standard deviation $\sigma$. Naturally, we can always go into a rotating reference frame with frequency $\bar{\omega}$ and therefore we can set $\bar{\omega}=0$ throughout our study without any loss of generality. We will use dimensionless units for all parameters and variables and fix $\sigma=1$ in this section. We point out that the oscillator coupling matrix $\mathbf{J} = (J_{n,m}) \in \mathbb{R}^{M\times M}$ has always at least one positive eigenvalue for all graphs considered throughout the paper which means that the trivial $\rho_n = 0$ solution is never stable. This means that the phases $\theta_n(t)$ are well defined at all times when calculating the dynamics of~\autoref{eq:SL}.

To understand the emergent coherence properties of the network it is useful to define a phase order parameter, commonly used in analysis of such networks~\cite{acebron_kuramoto_2005}, to capture the degree of phase coherence between the oscillators. For the complete graph it is written,
	\begin{equation}
	r_\text{complete} = \frac{1}{N}\left|\sum_{n=1}^{N}e^{i\theta_n}\right|.
	\label{eq:r_complete}
	\end{equation}
	If all the oscillators have the same phase $\theta_n = \theta_m$ then $r_\text{complete} = 1$, and in the limit of infinitely many uniform-randomly distributed phases on the interval $[0,2\pi)$ one has $r_\text{complete} = 0$. We have checked that the special case of $r_\text{complete} = 0$ where half of the oscillators are $\theta_n=0$ and other half $\theta_n = \pi$ (i.e., anti-phase ordering) does not appear in our results.
	
\begin{figure}[t]
		\centering
		\includegraphics[width=1\linewidth]{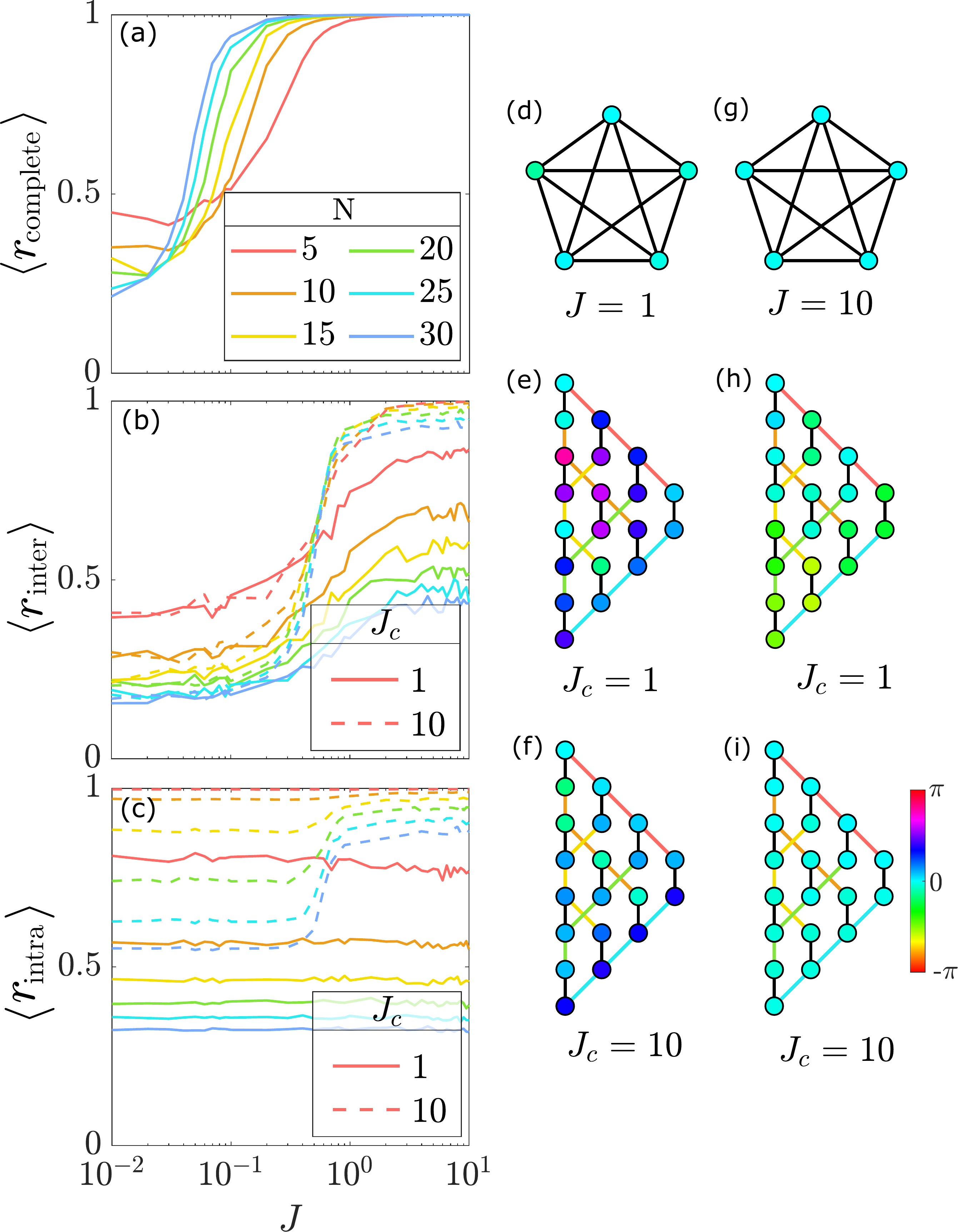}
		\caption{Coherence of (a) $\langle r_\text{complete} \rangle$ for uniform FM coupled complete graphs with coupling strength $J$, (b) $\langle r_\text{inter} \rangle$ and (c) $\langle r_\text{intra} \rangle$ for the corresponding triad graphs. All results are averaged over 160 instances of $\omega_n$ and random initial conditions. Plots (a-c) all share the same $x$-axis. Example steady state phases of $N=5$ oscillators in the (d,g) complete and triad structures with (e,h) $J_c=1$ and (f,i) $J_c = 10$ for a single instance. (d-f) $J = 1$ and (g-i) $J = 10$, where the phases are indicated by vertex color.
		}
		\label{fig:FM}
	\end{figure}

	We numerically integrate \autoref{eq:SL} from $t=0\to T\gg J^{-1},J_c^{-1}$ and calculate the average coherence $\langle r_\text{complete} \rangle$ at the final time $t = T$ over 160 random realizations of $\omega_n$ and initial conditions (i.e., Monte Carlo sampling). We then repeat our calculation over a range of coupling strengths $J$ [see \autoref{fig:FM}(a)] observing a gradual transition from an incoherent state to a coherent state with increasing coupling strength. This is reminiscent of the coherence bifurcation in Kuramoto networks~\cite{strogatz_kuramoto_2000,acebron_kuramoto_2005}. There, all phase oscillators are in an incoherent state $r_\text{complete}=0$ below some critical coupling strength $J_\text{crit}$ defined in the limit $N \to \infty$. As $J$ is increased through and above $J_\text{crit}$, the system reaches a partially synchronized state $r_\text{complete}>0$, where oscillators at the centre of $g(0)$ are synchronized while those at the tails of the distribution remain in an incoherent state such that the system is split in two dynamical groups \cite{strogatz_stability_1991}. As $J$ is increased further, more oscillators join the synchronized group until the entire system becomes coherent \cite{kuramoto_self-entrainment_1975,kuramoto_mutual_1984}, as we observe with the Stuart-Landau model. Finite size effects can also be clearly observed in \autoref{fig:FM}(a) when changing the number of oscillators $N$. Notably, larger coupling strengths are needed to synchronize smaller networks. We also point out that the presence of finite coherence values as $J \to 0$ arises from these finite size effects which gradually decrease as $N$ becomes larger.
	
	We now move onto the triad graph. Similar to \autoref{eq:r_complete}, we can define coherence order parameters for the unembedded (averaged) phases of the triad graph,
	\begin{equation}
	r_\text{inter} =\frac{1}{N}\Bigg|\sum_{n=1}^{N} e^{i \bar{\theta}_n}\Bigg|,
	\label{eq:r_unemb}
	\end{equation}
	and for the average phase coherence within each chain of the triad,
	\begin{equation}
	r_\text{intra}=\frac{1}{N}
	\sum_{n=1}^N
	\underbrace{
		\frac{1}{N-1}\Bigg| \sum_{n' \in \text{chain}}^{N-1}e^{i\theta_{n,n'}}\Bigg|
	}_\text{$r^\text{intra}_n$}.
	\label{eq:r_intra}
	\end{equation}
	For more details, please see \autoref{fig:SL_coherence} in \autoref{app:A}. Just like for the complete graph, we average the coherences of the triad over 160 random samples of the dynamics denoted by $\langle . \rangle$.

	Considering two different orders of the scaling parameter $J_c=\{1, 10\}$, we investigate how the triad graph's coherence properties relate to the complete graph (going to lower and larger orders of $J_c$ did not qualitatively change the findings). Note that the frequencies $\omega_n$ are randomly drawn from $g(\omega)$ for all oscillators in the triad which represents experimental reality (i.e., we do not associate a single random frequency across each chain). As expected, when $J_c$ is small the inter- and intra-chain coherences in \autoref{fig:FM}(b,c) show poor coherence with weak dependence on $J$. When $J_c$ is large we instead observe a fast inter-chain coherence transition in \autoref{fig:FM}(b). This indicates that the triad graph managed to represent the embedded complete graph dynamics within $J/J_c<0.1$ which gives a figure of merit for the design requirements of possible triad graph platforms. Interestingly, in \autoref{fig:FM}(b) for large $J$ the coherence is smaller in large graphs. This can be attributed to the fact that the chains within the triad graph need themselves to be coherent, $r_\text{intra} \approx 1$, in order to represent the embedded complete graph. But longer chains struggle more to settle on a phase and achieve good coherence as can be seen in \autoref{fig:FM}(c), such that the larger graphs in general require a stronger FM coupling strength in order to build coherence. 
	
As both $J$ and $J_c$ increase, both inter and intra-chain coherences converge to unity and all oscillators synchronize. In \autoref{fig:FM}(d-i) we show example simulations of varying coupling strength where the vertex colors represent the steady state oscillator phases. 
The effect of looping the chains like shown in \autoref{fig:triad_schem}(d) is found to have little effect on the coherence properties of the system (see \autoref{fig:looped}).

	\begin{figure}
		\centering
		\includegraphics[width=0.95\linewidth]{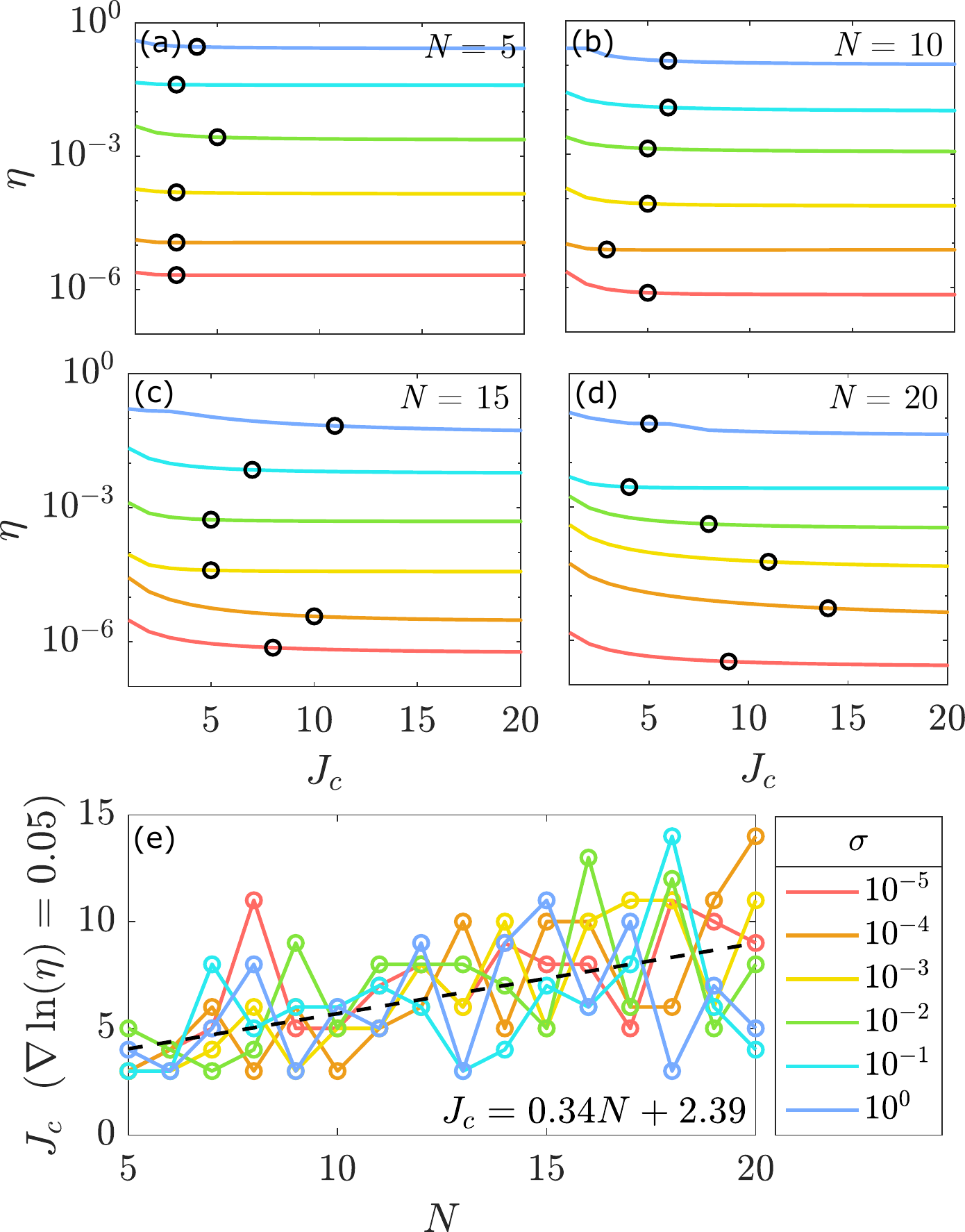}
		\caption{RMSE between the complete and triad graph phase dynamics for (a) $N = 5$, (b) $N = 10$, (c) $N=15$ and (d) $N=20$ oscillators over a range of $J_c$ and $\sigma$. 
		The black circles indicate the value of $J_c$ where the gradient of $\ln( \eta)$ surpasses 0.05. These points are collated in (e) and are plot as a function of $N$, where a linear fit is shown by a black dashed line following $J_c = 0.34N + 2.39$.}
		\label{fig:eta}
	\end{figure}

To better understand the quality of the triad graph as a simulator for the phase dynamics of the complete graph we calculate the root mean square error
(RMSE) between the phases in the complete graph $\theta_n$ and their corresponding representations in the triad graph $\bar{\theta}_n$ as a function of $J_c$. For brevity, we use the same set of randomly sampled detunings $\omega_n$ with strength $\sigma$, and same random initial conditions $\psi_n(t=0)$ across all calculations in order to quantify the similarity between the complete and the triad graph. We define a gauge-invariant RMSE measure as
\begin{equation}
\begin{split}
    \eta = \frac{1}{N(N-1)T}\Biggl[ \sum_{n<m} &\sum_{t=1}^T \Big[\cos\left(\theta_{nm}(t)) -  \cos(\bar{\theta}_{nm}(t)\right)\Big]^2 \\
    + &\Big[\sin\left(\theta_{nm}(t)) - \sin(\bar{\theta}_{nm}(t)\right)\Big]^2\Biggr]^\frac{1}{2}
\end{split}
\end{equation}
where $\theta_{nm} = \theta_n - \theta_m$ for the complete graph phases and $\bar{\theta}_{nm} = \bar{\theta}_n - \bar{\theta}_m$ for the representative triad graph phases. The RMSE is calculated only for the last 200 timesteps of each simulation in order to skip any transient effects that might be strongly dissimilar at early times between the complete and the triad graph. 

In \autoref{fig:eta}(a-d), we plot the RMSE with $N$ = 5, 10, 15 and 20 respectively, 
which show that the RMSE converges towards a minimum with increasing $J_c$, where we have marked with black circles the values of $J_c$ where the gradient of $\ln( \eta)$ surpasses 0.05.

The corresponding $J_c$ values where $\nabla \ln(\eta) = 0.05$ are collated in \autoref{fig:eta}(e), showing that the optimum value of $J_c$ increases with the number of oscillators and, interestingly, appears to be independent of the strength of detuning $\sigma$. We apply a linear fit to this data, shown by the black dashed line, with $J_c = 0.34N + 2.39$ which gives an estimation of what value of intra-chain coupling strength in the triad $J_c$ is needed in order to simulate the dynamics of the complete graph.

\subsection{XY Energy Minimization} \label{sec.XY}
In this section we investigate the feasibility in using the minor embedding technique on complete graphs of Stuart-Landau oscillators to optimize the XY Hamiltonian. We compare the unembedded and embedded complete graph XY energies extracted from the triad structure [\autoref{eq:unembedded},\ref{eq:embedded}] to the ground state of the XY Hamiltonian, as calculated using the basin hopping optimization method \cite{Wales_BasinHopping_1997}. Here we measure the performance of the Stuart-Landau system through the normalized difference in the extracted XY energy from solving the dynamics of the triad graph of Stuart-Landau oscillators $E_{SL}$ compared to the complete graph XY energy solved using the basin hopping method $E_{BH}$,
\begin{equation}
\text{Error} = \frac{E_{BH} - E_{SL}}{E_{BH}}.
\label{eq:error}
\end{equation}
\begin{figure}[t]
		\centering
		\includegraphics[width=1\linewidth]{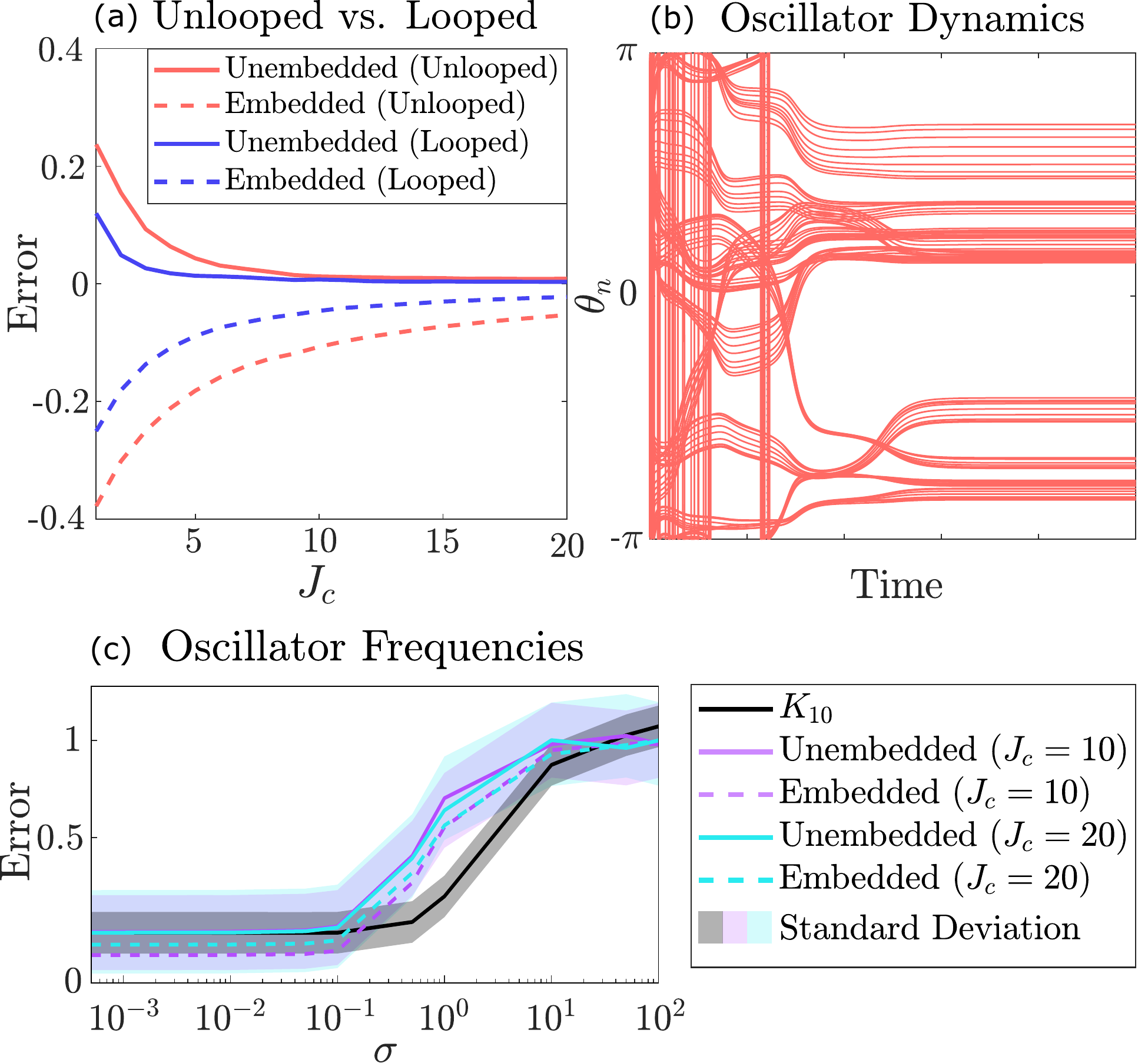}
		\caption{(a) Error between the triad embedded Stuart-Landau model and the basin hopping method in minimising the XY Hamiltonian on a randomly connected complete graph with $N=10$ vertices and $\sigma = 0$. (b) Example phase dynamics of the triad embedded oscillators for a single random graph with $J_c=10$, indicating clustering of the phases in different chains. (c) Performance of the system for increasing strength in oscillator frequencies $\omega_n$ for unlooped triad graphs with $J_c=10$ (purple), $J_c = 20$ (cyan) and the complete $K_{10}$ graph (black) for reference, with translucent surfaces to represent the standard deviation in error of the unembedded energies. Results in (a,c) are averaged over 160 random graphs and initial conditions and, in the case of (c), for random oscillator energies $\omega_n$.}	
		\label{fig:N10_randomXY}
\end{figure}
The basin hopping algorithm is time consuming but a highly accurate optimization algorithm that serves as a good reference for the energies found by the solving the dynamics of triad embedded Stuart-Landau networks. Previous studies on complete graphs (i.e., no embedding techniques) have shown very good performance from numerically solving the dynamics of Stuart-Landau networks as compared to the basin hopping algorithm~\cite{Kalinin_NJP2018}. For completeness, we provide evidence in \autoref{app:B} that the Stuart-Landau network indeed outperforms commercial optimizers, and also direct integration of the Kuramoto model, in finding low energy solutions of the XY Hamiltonian. Throughout this section, the performance is obtained by averaging over 160 unique complete graphs with weights randomly selected from a uniform distribution $J_{n,m} \in [-1,1]$ (there is no qualitative difference in using a normal distribution of same variance). 

To find the optimum embedding parameter $J_c$, we first consider a graph size of $N=10$ in \autoref{fig:N10_randomXY}. In \autoref{fig:N10_randomXY}(a) we show the system performance for $\sigma=0$ and scanning across the embedding parameter $J_c$. The error of the extracted XY energies reduces as $J_c$ is increased, where the unembedded XY energy converges to zero faster than the embedded energy, with a reduction in error for both methods when the triad chains are looped. This result is to be expected, as larger values of $J_c$ reduce the distribution of oscillator amplitude and phase across each triad chain, achieving a better representation of the complete graph in corroboration with \autoref{fig:FM}(b). 
	This is seen in \autoref{fig:N10_randomXY}(b) where the phase dynamics $\theta_n(t)$ of the $N(N-1) = 90$ triad graph oscillators split into 10 moving paths, with each path representing a different triad chain. 

Interestingly, looping the triad chains in the considered case of $N=10$ oscillators is found to reduce the phase variation within each chain (see \autoref{fig:looped}), achieving lower error (see \autoref{fig:N10_randomXY}). However, for larger $N$ the error starts increasing (see \autoref{fig:randomXY} in the Appendix) which we discuss in more detail at a later stage. We point out that the negative error for the embedded XY energy stems from the fact that $\text{min}[H_{XY}^\text{emb}] < \text{min}[H_{XY}]$ when $J_c \to 0$. In this limit, the triad graph breaks into $N$ pair-coupled oscillators (only black edges remain) and the sum of their XY energies is trivially minimized by simply setting the relative phase in each pair to $\Delta \theta_{n,m} = 0,\pi$ for positive and negative couplings, respectively.

In \autoref{fig:N10_randomXY}(c) we show the performance of Stuart-Landau networks in dynamically finding a good XY energy when including random oscillator frequencies $\omega_n$, normally distributed with standard deviation $\sigma$. Here, we focus on unlooped triad graphs with $J_c = 10$ and $J_c =20$. As an additional reference, we also plot the performance of the Stuart-Landau oscillators in the original complete $K_{10}$ graph (black line). As expected, the frequencies $\omega_n$ lead to desynchronization and fast increase in error with increasing $\sigma$, and the network steady state is lost with no well-defined phase relation that can enter into the XY Hamiltonian. This increase in error in the Stuart-Landau networks occurs for $\sigma \gtrsim \frac{1}{10}$ which is in qualitative agreement with coherence transition in \autoref{fig:FM}(b). 
 
	\begin{figure}[t]
	\centering
	\includegraphics[width=1\linewidth]{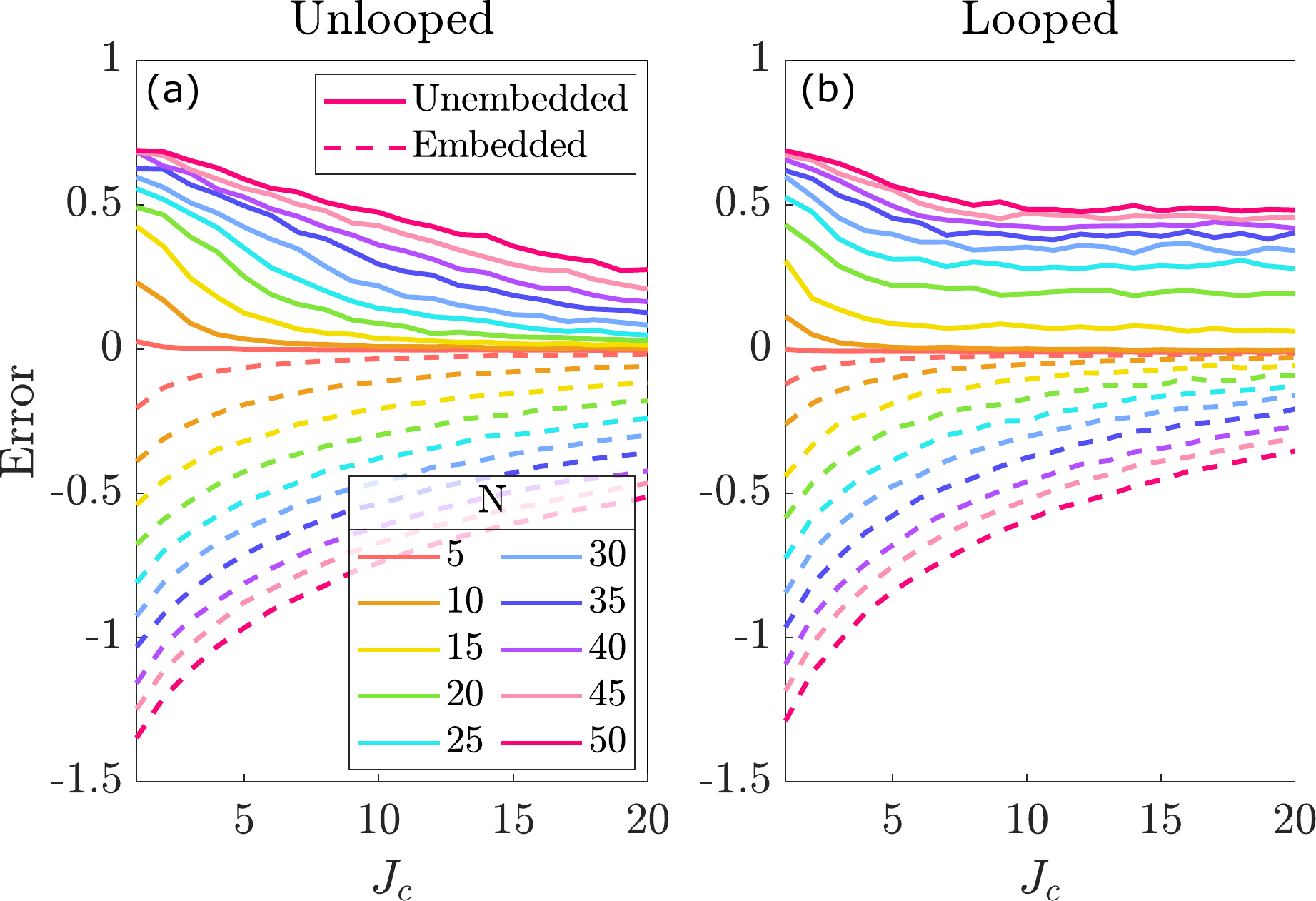}
	\caption{XY energy error from the triad structure of Stuart-Landau oscillators compared to the complete graph XY energy solved using the basin hopping method for (a) unlooped and (b) looped triad chains with $N = 5$ to $50$, averaged over 160 random graphs and $\sigma = 0$. }
	\label{fig:randomXY}
\end{figure}  	

Lastly, we show the performance of the triad graph as a function of both the problem (i.e., network) size $N$ and $J_c$ in \autoref{fig:randomXY}. As before, the error converges to zero with increasing $J_c$ with better performance for the unembedded XY energy. As expected, larger networks struggle to find the correct (global) XY energy minimum and instead stabilize into the growing number of local minima. Interstingly, in \autoref{fig:randomXY}(b), when the chains are looped the unembedded XY energy error plateaus such that beyond $J_c=10$, the embedded energy does not noticeably change and the benefit of looping the chains is only apparent for $N\leq10$. This puzzling behaviour implies that looping the chains in the triad graph has generated a new family of stable attractors which do not correlate with the minima of the XY Hamiltonian. Such attractors could belong to twisted states in the chains which appear in sparse networks~\cite{Townsend_Chaos2020}, which we explore in the following section.

\subsection{Twisted States}
Twisted states in oscillatory networks are characterized by a winding number, $\ell$, denoting integer multiples of 2$\pi$ phase winding about a network of FM coupled oscillators~\cite{Townsend_Chaos2020}. Typically, these states appear in looped networks (where the first and last oscillators in a chain are connected) with sparse connectivity up to the first few nearest neighbors. Although an interesting nontrivial state, this family of attractors is however detrimental to the performance of the looped triad graph to find good XY energy minima. For the triad to best represent the complete graph, the FM chains in the triad should have all sites following same phase dynamics (or as similar as possible) in order to represent the dynamics of a single oscillator in the complete graph. Appearance of a vortex phase gradient in the looped triad chains is clearly detrimental to this requirement.

Networks of $N$ identical Kuramoto oscillators have recently been studied with unit FM coupling to $\mu(N-1)$ nearest neighbors \cite{Townsend_Chaos2020}, where $\mu > 0.75$ guarantees that all oscillators will converge to a synchronized in-phase configuration with $\ell = 0$ . Below this threshold however, the loop of oscillators exhibit a constant relative phase difference between nearest neighbors of $2\pi\ell/N$, where $\ell\neq 0$. These solutions are known as \textit{twisted states}.

	\begin{figure}
	\centering
	\includegraphics[width=1\linewidth]{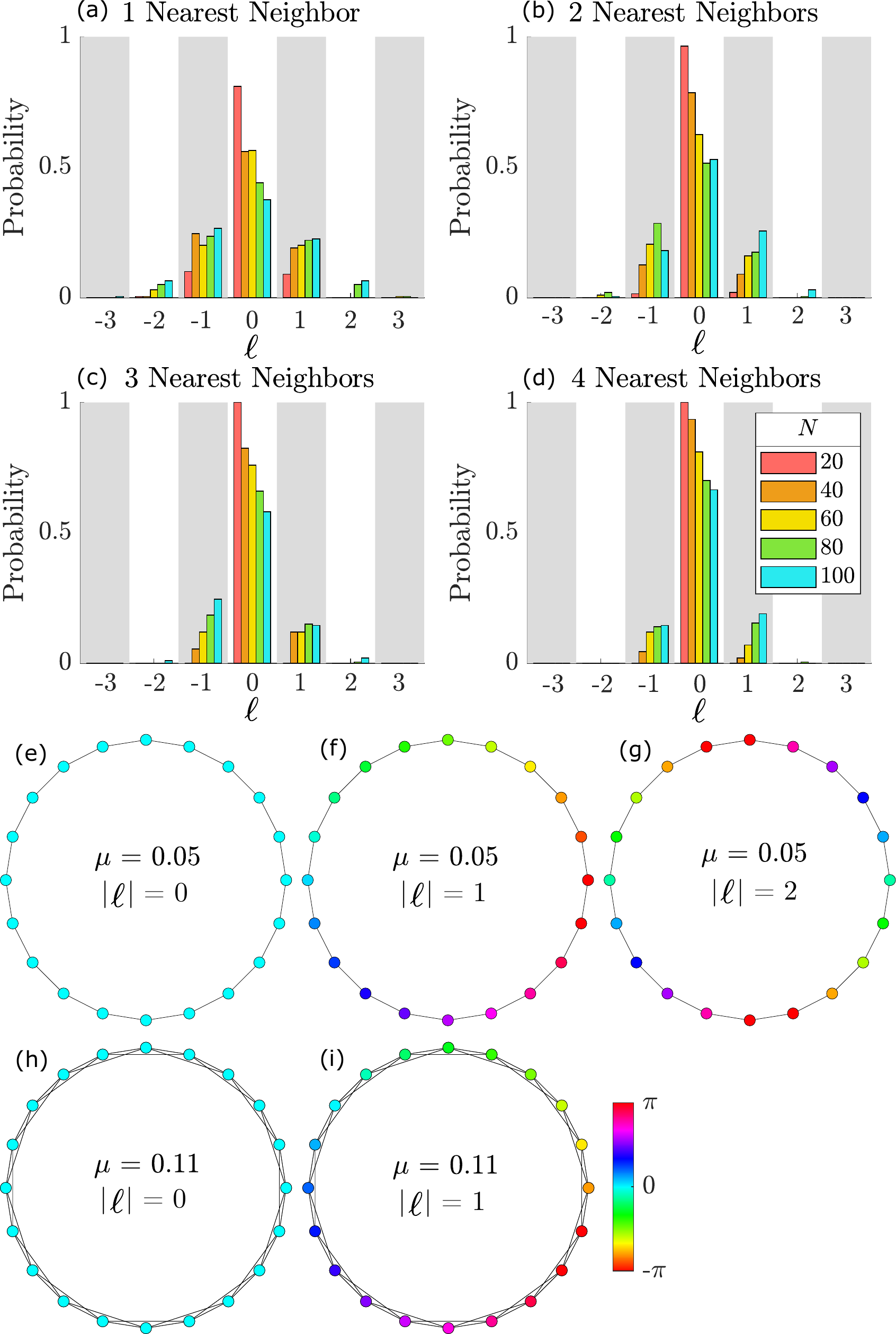}
	\caption{Distribution in winding number $\ell$ extracted from a continuous chain of Stuart-Landau oscillators with (a-d) 1-4 nearest neighbor couplings with unitary strength ($J = 1$) respectively for $N =$~20, 40, 60, 80 and 100 shown in red, orange, yellow, green and blue bars respectively, over 1000 unique realizations. For $N =$~20, an example of each stable phase solution is shown in graph format with (e-g) 1 and (h,i) 2 nearest neighbor connections, showing all observed values of $|\ell|$ over 1000 realizations.} 
	\label{fig:twistedStates}
\end{figure}

To explore the occurrence of twisted states in the Stuart-Landau model, we create networks of $N$ identical oscillators ($\sigma$ = 0) in loops (just like a single looped chain of a triad graph) with unitary FM coupling and plot the distribution of different winding numbers $\ell$ that appear in the system over 1000 unique realizations as a function of nearest neighbor connectivity and varying $N$ in \autoref{fig:twistedStates}(a-d). Remarkably, we find that twisted states appear frequently in our simulations. For larger $N$, the spread in $\ell$ is greater, but reduces quickly as the number of nearest neighbor connections increases. When $N = 20$ for example, we observe $\vert \ell \vert = 0,1,2$ for first nearest neighbor coupling, but this range decreases to $\vert \ell \vert = 0, 1$ when second nearest neighbor coupling is introduced, as depicted in \autoref{fig:twistedStates}(e-i) where the oscillators (circles) have unit FM coupling following the black graph edges and their steady-state phases are shown by their color. In this analysis, $|\ell| = 2$ occurred in 0.1\% of the cases for $N = 20$ with first nearest neighbor connectivity, so it is very unlikely to observe a phase winding greater than $2\pi$ in a smaller chain. This is the likely reason that looped-chain triad graphs do not converge to zero energy error in \autoref{fig:randomXY}(b), but instead plateau to a larger error that grows in value with increasing $N$. As non-zero winding numbers in Stuart-Landau chains grows in probability with $N$, we emphasize that it is important to sample the system over many realizations starting with different random initial conditions in order to increase the chances of finding an XY energy closest to the XY ground state.\\

\subsection{Dynamic Pumping}\label{subsec:feedback}
\begin{figure}
	\centering
	\includegraphics[width=1\linewidth]{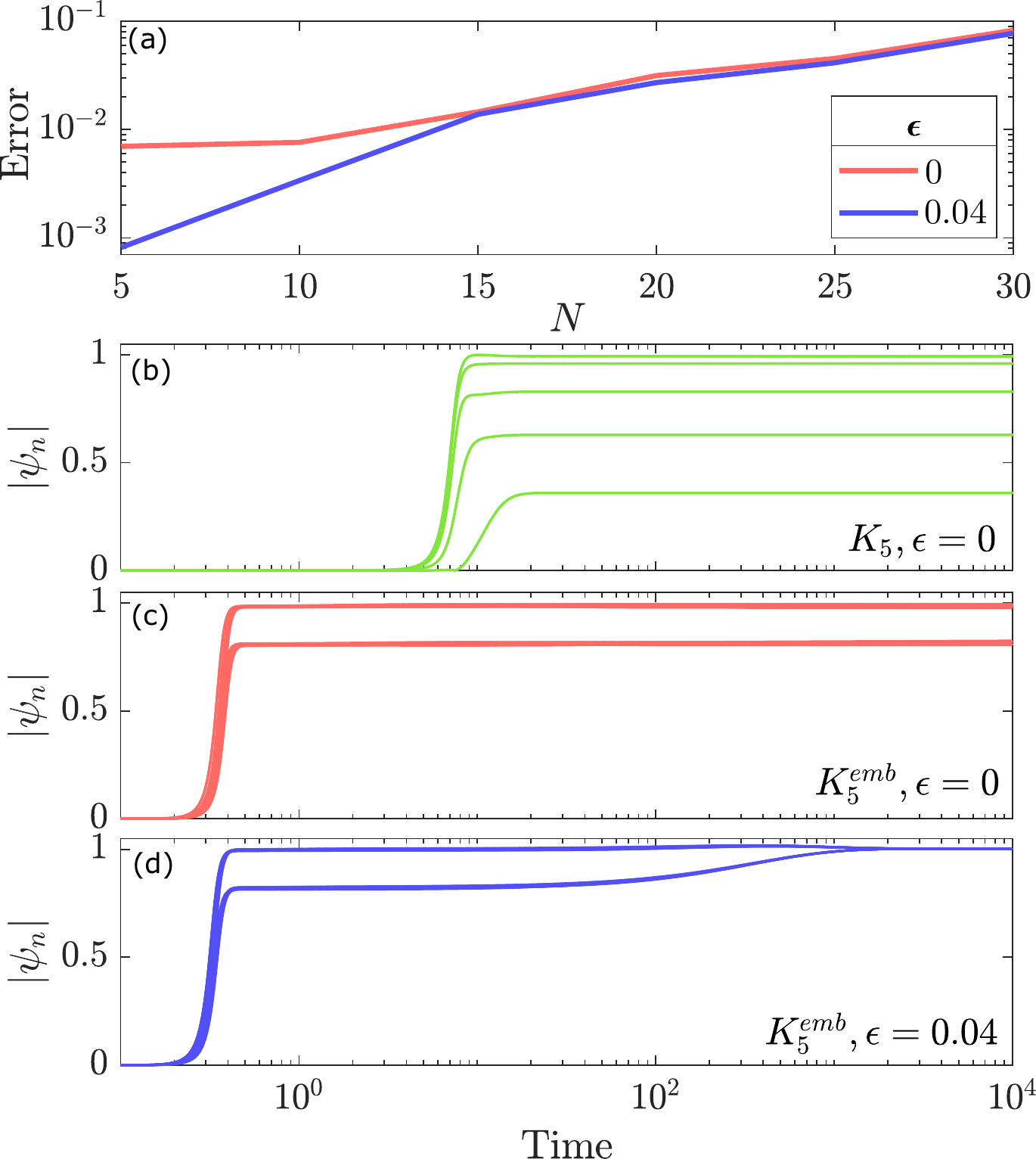}
	\caption{(a) Unembedded energy error of randomly connected triad graphs with (red) $\epsilon=0$ and (blue) $\epsilon=0.04$ with $J_c=20$ over a range of $N$, averaged over 100 unique random graphs realizations. Oscillator amplitude dynamics for (b) a randomly connected $K_5$ graph with $\epsilon = 0$, and the corresponding $J_c = 20$ triad graph with (c) $\epsilon = 0$ and (d) $\epsilon = 0.04$, with amplitudes normalized to $\max|\psi_n|$ and $\rho_t = \max|\psi_n|$. The same $x$-scale is used for panels (b-d).	}
	\label{fig:feedback}
\end{figure}

The presence of random couplings induces an amplitude inhomogeneity in the Stuart-Landau network $\rho_n~\neq~\rho_m$, meaning its fixed points will, in general, not coincide exactly to the minima of the XY model. As a possible improvement to achieve amplitude homogeneity~\cite{Kalinin_NJP2018, Inui_CommPhys2022}, we investigate the addition of a dynamic pumping mechanism that feeds back the amplitude of each oscillator at each step in numerical integration in order to adjust the gain of each node respectively. To explore this feedback mechanism, we adjust \autoref{eq:SL} slightly to follow:
\begin{equation}
	\frac{d\psi_n}{dt} = [P_n(t) - i\omega_n - |\psi_n|^2]\psi_n + \sum_{m=1}^N J_{n,m}\psi_m.
	\label{eq:SL_feedback}
\end{equation}
Here, $P_n(t)$ is the pumping rate of oscillator $\psi_n$ where initially all oscillators are injected equally with $P_n(0) = 0$ for all $n$. In terms of exciton-polariton condensates~\cite{berloff_realizing_2017, Tao_NatMat2022}, the pumps represent an arrangement of spatially modulated nonresonant excitation beams. After the initial pumping in the first time step of integration, we apply a feedback mechanism to balance the occupation of each oscillator, following:
\begin{equation}
	\frac{dP_n}{dt} = \epsilon \left[\rho_{t} - \rho_n(t)\right]
	\label{eq:SL_eps}
\end{equation}
where $\epsilon$ controls the rate of change of $P_n$, $\rho_{t}$ is the target amplitude of the oscillators. We numerically integrate Equations~\ref{eq:SL_feedback} and \ref{eq:SL_eps} for 100 unique randomly-connected complete graphs and corresponding triad graphs with varying $N$, $J_c=20$, and for $\epsilon = 0$ and $\epsilon = 0.04$. The resulting performance, and amplitude dynamics for a single instance of the network is shown in \autoref{fig:feedback}.

We do indeed see a considerable drop in error for smaller networks when the feedback is present. However, for larger networks the error is dominated by the approximative nature of the embedding procedure and the role of feedback becomes less important. It could be possible to design a more complex feedback procedure which not only eliminates amplitude inhomogeneity but also helps balance the phases in each chain of the triad graph, but this is beyond the scope of the current study. Note, due to the large intra-chain couplings $J_c$ in the triad graph, the oscillator amplitudes reach saturation an order of magnitude sooner than for the equivalent complete graph.

	\section{Conclusions}
	We have demonstrated that the dynamics of an random all-to-all coupled Stuart-Landau oscillator network can be approximated using a minor embedding technique, regularly applied in the design of quantum computing platforms. Here, a complete (dense) graph is embedded into a sparse triad graph defined by a single embedding parameter $J_c$. We show that the steady-state phases in the embedded Stuart-Landau oscillator network correspond to good approximation of the optimal solutions in the corresponding graph XY Hamiltonian, achieving good performance by simply adjusting its embedding parameter. The results are compared against the standard complete graph Stuart-Landau oscillator dynamics, and the basin hopping method.
	
	  The convergence of the minor embedded graph dynamics to that of the complete graph offers up the triad structure as a potential testbed for mapping out dense graph problems to continuous-phase coupled oscillator systems where fully-controllable all-to-all couplings are not practicable (such as polariton condensates, photonic condensates, and coupled laser arrays). This opens perspectives on designing analogue computing hardware aimed at heuristically solving dense graphs problems (such as optimising the XY Hamiltonian) across a wide range of platforms in a similar spirit to quantum computing platforms. Optical systems might hold a particularly strong promise in this regards since their GHz operation rates can provide rapid sampling of the objective function energy landscape that can serve as good trial points for more sophisticated optimizers. 
	  
	  \appendix
	  \section{Coherence}\label{app:A}
	\subsection{Definitions}
	
\begin{figure*}
		\centering
		
		\begin{tabular}{p{6cm}p{6cm}p{6cm}}
			
			\adjincludegraphics[width= 3.7cm, valign = c]{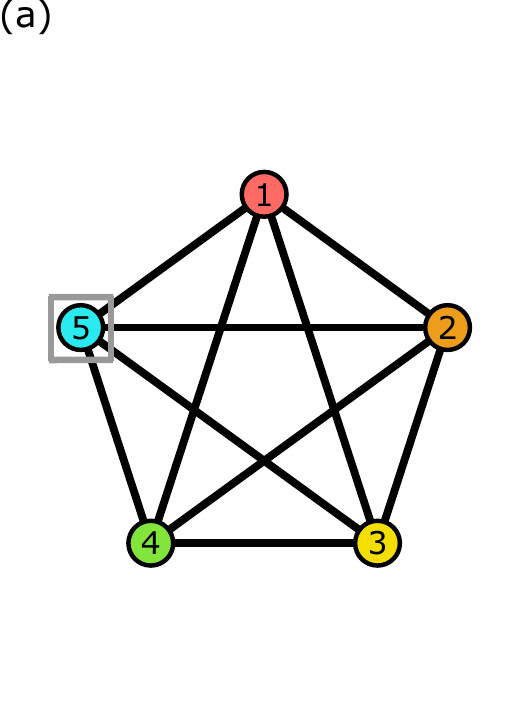}
			&
			
			\adjincludegraphics[width= 4.5cm, valign = c]{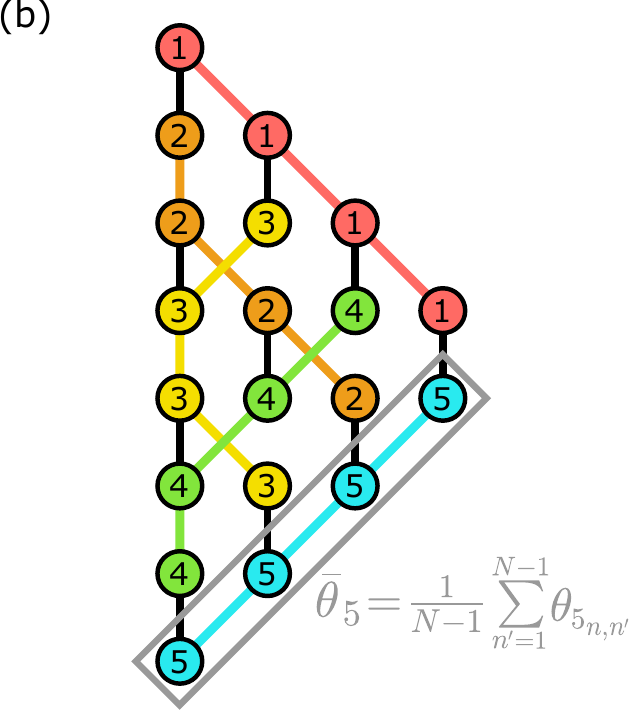}
			&
			
			\adjincludegraphics[width= 3.8cm, valign = c]{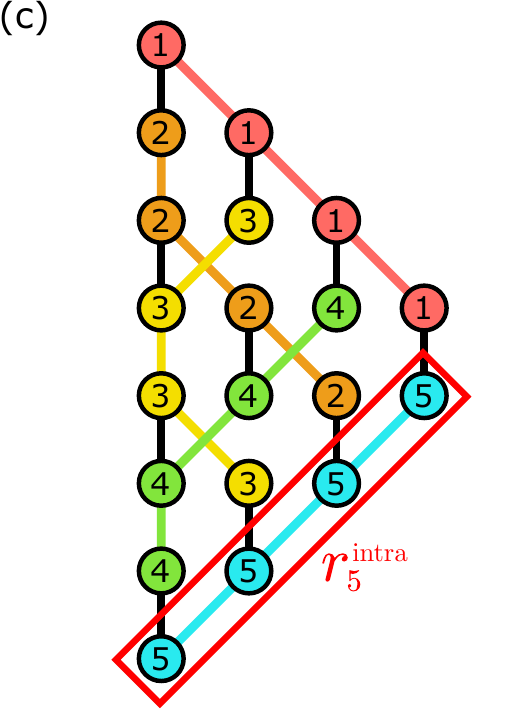}\\
			
			&&\\
			\(\displaystyle r_\text{complete}=\frac{1}{N}\Bigg|\sum_{n=1}^{N} e^{i\theta_n}\Bigg|\)
			&
			\(\displaystyle r_\text{inter}=\frac{1}{N}\Bigg|\sum_{n=1}^{N} e^{i \bar{\theta}_{n}}\Bigg|\)
			&
			\(\displaystyle r_\text{intra}=\frac{1}{N}
			\sum_{n=1}^N
			\underbrace{
				\frac{1}{N-1}\Bigg| \sum_{n' \in \text{chain}}^{N-1}e^{i\theta_{n,n'}}\Bigg|
			}_\text{$r^\text{intra}_n$}.\)\\
		\end{tabular}
		\caption{(a) Coherence extracted over all oscillators in the complete graph, (b) the unembedded triad coherence between the $N$ average chain phases, plus (c) the average chain coherence. All schematics represent the $K_5$ and $K_5^{emb}$ graphs with the grey boxes in (a,b)  and red box in (c) representing the extraction of the average phase and coherence across chain 5 respectively.}
		\label{fig:SL_coherence}
		\end{figure*}

	In the main text we extract  the coherence across all oscillators in the complete graph $r_\text{complete}$, the coherence across the average phases of the chains $r_\text{inter}$ and the average coherence across each chain $r_\text{intra}$. All three parameters are defined alongside a schematic in \autoref{fig:SL_coherence}. Here, $\theta_n$ is the phase of oscillator $n$ in the complete graph, $\theta^\text{intra}_{n_{n'}}$ is the phase of the $n'^{th}$ oscillator in triad chain $n$ and $\bar{\theta}_{n}$ is the average phase over the $N-1$ oscillators in chain $n$.

    \subsection{Looping Triad Chains}
\begin{figure} 
		\centering
		\includegraphics[width=1\linewidth]{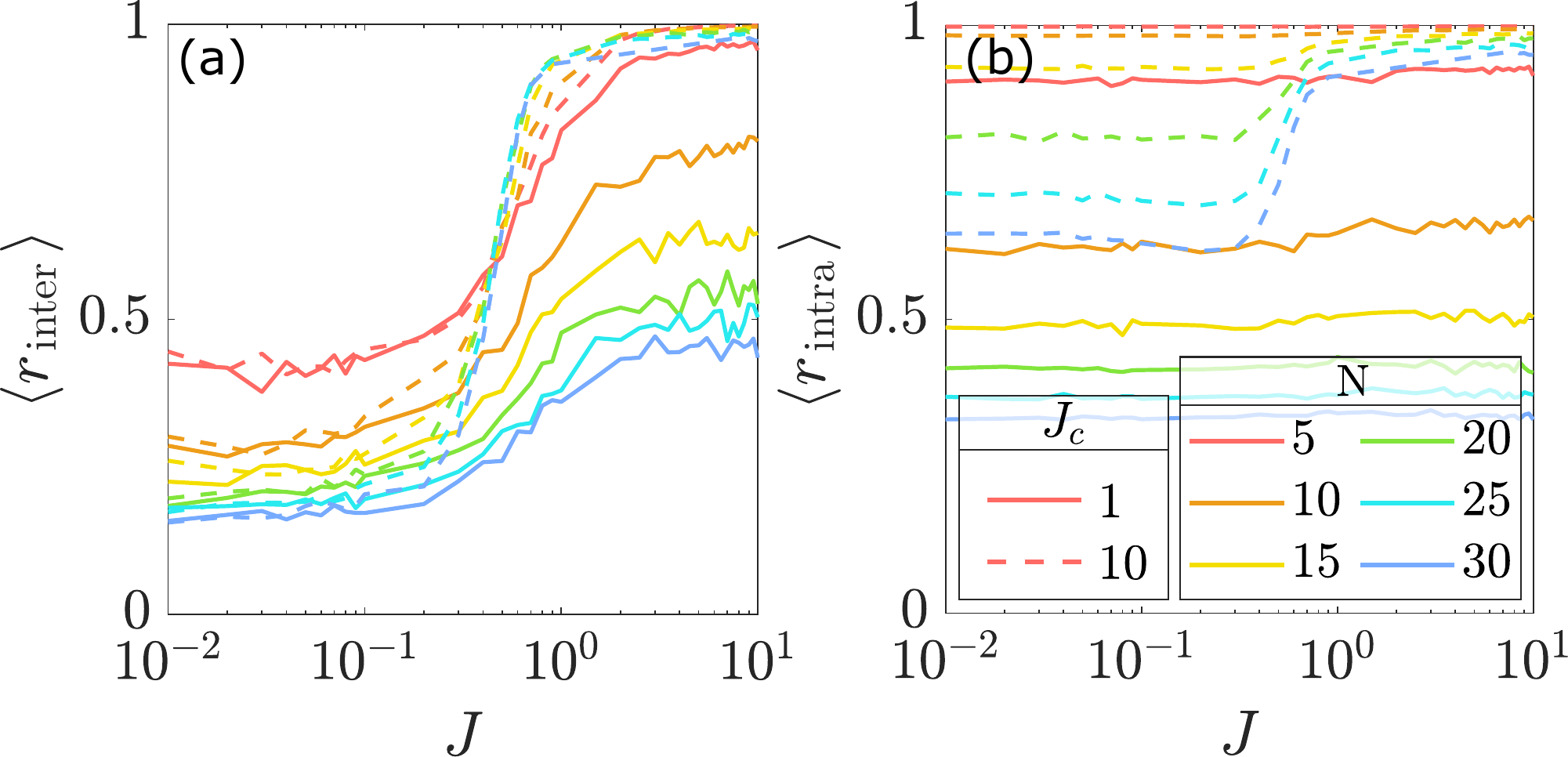}
		\caption{ Coherence (a) $\langle r_\text{inter}\rangle$ and (b) $\langle r_\text{intra}\rangle$ of triad graphs corresponding to uniform FM complete graphs with looped chains and $J_c = 1$ and $10$. All scanned over a range of coupling strengths $J$ and averaged over 160 instances with standard deviation in oscillator frequencies $\sigma = 1$.}
		\label{fig:looped}
	\end{figure}

	In addition to the unlooped triad representation of the uniform ferromagnetic (FM) complete graph in the main text, we also consider the effect of looping the triad chains on the coherence of the system under the same conditions as the unlooped case [\ref{fig:looped}]. When the minor embedded chains are looped, all oscillators are symmetrically coupled and thus the triad graph coherence is higher for the same coupling strength compared to the unlooped equivalent, as the edge effects of the unlooped chain coherence are eliminated. For $J_c = 1$, $r_\text{inter}$ and $r_\text{intra}$ remain at a low coherence as the encoded complete graph weights are not dominant in the system. For $J_c=10$, there is a coherence build up with increasing $J$, indicating that the triad coherence dynamics represent the complete graph for large $J_c$.
	
	\section{Benchmarking Stuart-Landau networks in finding low energy XY solutions}\label{app:B}
	\subsection{Stuart-Landau networks against global optimizers using trust-region methods}
	Here we will benchmark the performance of dense Stuart-Landau (SL) networks against commercially available global optimizers. We have applied the {\it Basin Hopping} (BH) algorithm from the Python SciPy library as a benchmark in the main manuscript. However, we point out that we are benchmarking dense graphs meaning that a $N = 50$ vertex graph has $N(N-1) = 2450$ weighted edges to be optimized. The reason we limit ourselves to $N=50$ vertices in this study is because running the BH algorithm with 1000 iterations (for good convergence) for larger graphs on a single 2.6 GHz Intel Sandybridge processor (components of the IRIDIS 4 supercomputer at University of Southampton) exceeded 100 hours. 
		\begin{figure*}
		\centering
		\includegraphics[width=1\textwidth]{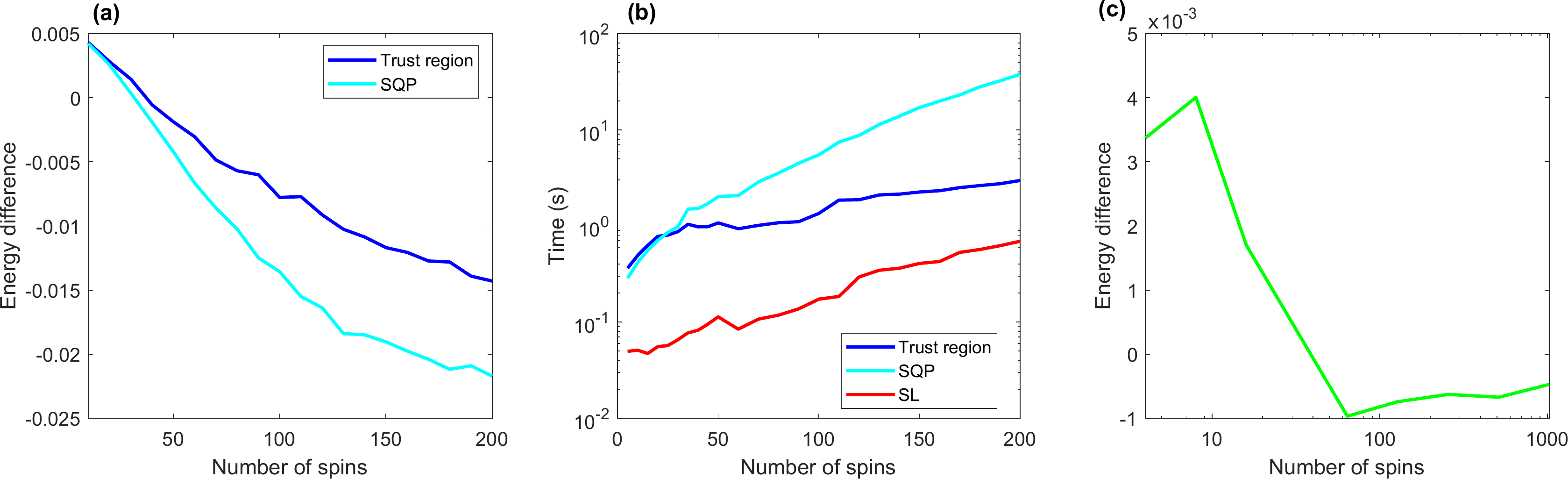}
		\caption{(a) Average energy difference between the GS algorithm and the SL model according to Eq.~\eqref{eq.1} with the GS algorithm using two different optimizer approaches, trust region and SQP. The SL model outperforms the GS algorithm around $N \approx 30$ spins. (b) Time taken for different methods to arrive at an answer for the XY energy. (c) Energy difference between the KM and the SL model [i.e., $E_\text{GS}$ is replaced with $E_\text{KM}$ in Eq.~\eqref{eq.1}]. The SL model outperforms the KM model in finding low energy values to the XY Hamiltonian at around $N \approx 30$.}
		\label{fig4}
	\end{figure*}
	
	As a result, to investigate graphs of $N>50$, instead of using the BH algorithm we have decided to use the speedier and commercially available global optimizer GlobalSearch~\footnote{\url{https://uk.mathworks.com/help/gads/how-globalsearch-and-multistart-work.html}} (GS) from the Global Optimization Toolbox~\footnote{\url{https://uk.mathworks.com/help/gads/}} of Matlab\textsuperscript{TM}. This algorithm uses a scatter search method to generate feasible trial points which are then evaluated using a chosen optimizer method to find multiple local minima. The GS algorithm then iteratively analyses points that converge using a score function which updates on-the-fly and rejects those points that are unlikely to improve the best minimum found so far. We have decided to use two well known optimization methods for the GS algorithm. \textbf{(1)} The {\it trust region} method since we can easily compute both the gradient and the Hessian matrix of the XY Hamiltonian (i.e., the objective function) making it quite fast and accurate and thus an appropriate choice. \textbf{(2)} The {\it sequential quadratic programming} (SQP) gradient descent method since it also benefits from knowing the gradient of the objective function. Our search region is bounded on $\theta_n \in [-4,4]$ which is taken larger than the periodic range $[-\pi,\pi]$ in order to more efficiently find minima that might be close to values around $\theta_n = \pm \pi$. All other options of the GS algorithm were set to default as they did not considerably improve the efficiency of the optimizer.
	
	Just like in the main manuscript, we quantify the performance between the SL model and the GS algorithm using the ratio of their energy difference,
	\begin{equation}
	\text{Energy Difference} = \frac{E_\text{GS} - E_\text{SL}}{2 E_\text{GS}}.
	\label{eq.1}
	\end{equation}
	The factor $1/2$ is added so that if $E_\text{SL} = - E_\text{GS}$ then the difference is exactly unity. Our results are presented in~\ref{fig4}(a) where we have averaged over 1000 random dense graphs with weights sampled from the interval $J_{n,m} \in [-1,1]$ going up to $N=200$ spins. Amazingly, even after supplying knowledge of the gradient and the Hessian of the objective function ($H_{XY}$ in main manuscript) the SL model starts outperforming the GS algorithm around $N \approx 30$ spins. We also show in~\ref{fig4}(b) the time taken to iterate the SL model using an explicit Runge-Kutta (4,5) method until it converged to a steady state (red line), and the time it took the trust region (blue line) and gradient descent (cyan line) methods in the GS algorithm to provide a solution. The results evidence the considerably better efficiency in iterating the SL model than applying the GS algorithm in finding a low energy value to the XY Hamiltonian.

	\subsection{Stuart-Landau versus Kuramoto networks}
	
	We additionally investigate the performance of the SL network at minimising the XY Hamiltonian in comparison to the fixed point solutions of the Kuramoto (KM) model. As mentioned in the manuscript, relating the steady states of the SL model with the minima of the XY Hamiltonian is a heuristic approach since, in general, oscillatory systems like lasers and driven-dissipative condensates have freely evolving amplitudes and therefore it demands investigation on how well such heuristic systems will perform in approximating the XY ground state energy. To backup this point, we show in~\ref{fig4}(c) the average energy difference between the KM and the SL model in minimising the XY Hamiltonian over 1000 random dense graphs. Amazingly, around $N \approx 30$ spins (oscillators) the SL model starts outperforming the KM model. This means that the SL oscillators are much more efficient in exploring their state space during their transient growth phase from the initial ``vacuum state'' $|\psi_n(t=0)| \simeq 0$ (in the context of quantum annealing) rather than the KM oscillators which are always locked onto the unit circle. Therefore, SL oscillators can clearly outperform KM oscillators at minimising the XY Hamiltonian of densely connected oscillator networks. \\

	  \section*{Acknowledgements}
	  S.L.H., H.S., and P.G.L. acknowledge the support of the UK’s Engineering and Physical Sciences Research Council (grant EP/M025330/1 on Hybrid Polaritonics). H.S. and P.G.L. also acknowledge the European Union's Horizon 2020 program, through a FET Open Research and Innovation Action under the Grant Agreement No.~899141 (PoLLoC) and No.~964770 (TopoLight). H.S. acknowledges the Icelandic Research Fund (Rannis), Grant No.~217631-051. S.L.H. acknowledges the use of the IRIDIS High Performance Computing Facility and associated support services at the University of Southampton. 
	  
	  \section*{Data Availability} All data supporting this article is available on the University of Southampton's online repository (DOI: \href{https://doi.org/10.5258/SOTON/D2429}{10.5258/SOTON/D2429}).


%

\end{document}